\documentclass[twocolumn,twoside,slac]{revtex4}
\usepackage{graphicx}
\usepackage{fancyhdr}
\RequirePackage{url}
\begin{document}

\newcommand{\DZ}{D\ensuremath{\emptyset}}
\newcommand{\brackets}[1]{\ensuremath{<}{\tt #1}\ensuremath{>}}

\title{The StoreGate: a Data Model for the Atlas Software Architecture}

%

\author{P. Calafiura, C.G. Leggett, D.R. Quarrie}
\affiliation{Lawrence Berkeley National Lab, Berkeley, CA 94720, USA}
\author{H. Ma, S. Rajagopalan}
\affiliation{Brookhaven National Lab,  Upton, NY 11973-5000, USA}

\begin{abstract}
The Atlas collaboration at CERN\cite{ATLAS} has adopted the Gaudi software
architecture which belongs to the blackboard family: data
objects produced by knowledge sources (e.g. reconstruction modules)
are posted to a common in-memory data base from where other modules
can access them and produce new data objects.  The StoreGate has been
designed, based on the Atlas requirements and the experience of other
HENP systems such as Babar, CDF, CLEO, D0 and LHCB, to identify in a
simple and efficient fashion (collections of) data objects based on
their type and/or the modules which posted them to the Transient Data
Store (the blackboard). The developer also has the freedom to use her
preferred key class to uniquely identify a data object according to
any other criterion. Besides
this core functionality, the StoreGate provides the developers with a
powerful interface to handle in a coherent fashion persistable
references, object lifetimes, memory management and access control
policy for the data objects in the Store. It also provides a
Handle/Proxy mechanism to define and hide the cache fault mechanism:
upon request, a missing Data Object can be transparently created and
added to the Transient Store presumably retrieving it from a
persistent data-base, or even reconstructing it on demand. 
\end{abstract}

\maketitle

\thispagestyle{fancy}


\section{INTRODUCTION}

\subsection*{Data Objects and Algorithms}
 The Gaudi software architecture\cite{Gaudi} belongs to the blackboard
family\cite{blackboard}: data objects produced by knowledge modules (called Algorithms in Gaudi) are posted to a common ``in-memory
data base'' from where other modules can access them and produce new
data objects. 

This model greatly
reduces the coupling between knowledge modules containing the
algorithmic code for analysis and reconstruction, since one knowledge
module does not need anymore to know which specific module can produce
the information it needs nor which protocol it must use to obtain it
(the "interface explosion" problem described in component software
systems). Algorithmic code is known to be the least stable component
of software systems and the blackboard approach has been very
effective at reducing the impact of this instability, from the Zebra
system of the Fortran days to the InfoBus architecture for Java
components. The trade-off of the data/knowledge objects separation is
the need for knowledge objects to identify data objects to be posted
on or retrieved from the blackboard. It is crucial to develop a data
model optimized for the required access patterns and yet flexible
enough to accommodate the unexpected ones.

\subsection*{The Transient Data Store}

 The Transient Data Store (TDS) is the blackboard of the 
Gaudi architecture: a module creates a data object and post 
it to the TDS to allow other modules to access it\footnote
{
  to be precise the current TDS implements only a ``passive'' blackboard,
  since modules do not react to TDS events (e.g. executing after a data object
  is registered into the TDS)
}. 

 Once an object is posted on to the store, the TDS takes ownership
of it and manages its lifetime according to preset policies, removing,
 for example, a TrackCollection when a new event is read.
The TDS also manages the conversion of a data object from/to its persistent 
form and provides therefore an API to access data stored on persistent media.

\section{StoreGate Design and Functionality}
StoreGate (SG), in common with most other existing data models, is basically 
a dictionary of data objects which manages their memory and oversees
conversion to/from persistency. The SG design process has been
informal and iterative. We released early and often and used
developers feedback to adjust our initial design concept\footnote{
 which was in any case largely based on ideas which have worked in
existing data models}. The result may lack the coherency of a formal
top-down design but it follows a few principles which have proved to
be useful.

\subsection*{Work with User Types}
The success of the STL and of other public domain template libraries means that it has become vital to design an open system that can work
with generic types that export an interface, in particular the STL 
containers, rather than forcing data objects to import a common interface.
SG adapts its behavior to the functionality each data
object exports. The only SG-imposed constraint on a data object\footnote{
this does not mean that the data model, simulation and reconstruction
groups should not issue design guidelines to ensure that ATLAS data
objects behave consistently in terms of memory management and 
persistability} is to be an STL {\it Assignable} type\cite{STL}.

\subsection*{Avoid User-defined Keys}
 The disadvantage of the data/knowledge objects separation is the need for
knowledge objects to identify data objects to be posted on or retrieved
from the blackboard. It is crucial to develop a data model optimized for 
the required access patterns and yet flexible enough to accommodate the
unexpected ones.

 SG addresses this problem with a two-step approach: it defines a
natural identifier mechanism for data objects and it
transparently associates  to each
data object a default value of this identifier allowing developers to
register and retrieve data objects without having to identify them
explicitly.  

 The first component of the identifier is the data object
type. Experience shows that HEP developers
tend to group the objects they work on into collections. As a result the TDS will often contain a single instance
of a data object type (say a {\tt TrackCollection} or several closely related
ones (e.g. a TrackCollection for each component of the Inner
Detector). The SG retrieve interface covers these two use cases (see
Fig.~\ref{fig:SGAPI}).

Type-based identification is not always sufficient.
For example the TDS may contain several equivalent instances of a
TrackCollection produced by alternative tracking algorithms.
Therefore we need to add a second component to our identification
mechanism: the identifier of the Algorithm instance that produced the
data object we want\footnote{ 
 notice that we need to identify the instance rather than the
class. In an often quoted use case, clients may want to distinguish among
tracks reconstructed by the same tracking algorithm using different
jet cone sizes.}. In the spirit of working with user types, the
SG will allow developers to augment this history identifier
with a generic key type optimized for their access patterns.

\begin{figure*}
  \begin{verbatim}

    //record a TrackCollection
    TrackCollection* pTrackColl = myTrackMaker.make();
    StatusCode sc = record(pTrackColl, ``MyTrackCollection'');

    //get the default TrackCollection 
    const TrackCollection* pTrackColl; 
    sc=sg->retrieve(pTrackColl);  

    //get my special TrackCollection
    TrackCollection* pMyTrackColl;  //non-const access may be restricted
    sc=sg->retrieve(pMyTrackColl, ``MyTrackCollection'');  

    //access all track colls using a pair of STL forward iterators
    DataHandle<TrackCollection> beginTrackColls, endTrackColls;
    sc=sg->retrieve(beginTrackColls, endTrackColls); //get all TrackColls
  \end{verbatim}
  \caption{\label{fig:SGAPI} The basic StoreGate Data Access API}
\end{figure*}

\subsection*{Control Object Access and Creation}
The TDS is the main channel of communication among modules.  
A data object is often the result of a collaboration among several
modules. SG allows a module to use transparently a
data object created by an upstream module or read from disk.

A {\it Virtual Proxy}\cite{GoF} defines and hides
the cache-fault mechanism: upon request\footnote{
 Currently the proxy uses lazy instantiation (i.e. the object is created
 only when the handle is dereferenced).
}, a missing data object
instance can be transparently created and added to the TDS,
presumably retrieving it from a persistent data-base or, in
principle, even reconstructing it on demand.  

To ensure reproducibility of data processing, a data object
should not be modified after it has been published to
the store, we use the same proxy scheme to enforce an ``almost const''
access policy: modules downstream of the publisher are only allowed to retrieve
a constant iterator to the published object.

\subsection*{Support Inter-object Relationships}
SG supports uni-directional inter-objects relationships, or
links, and will support bi-directional links in the
future. A link is a persistable pointer. If the linked object is a
data object then the proxy scheme described above is also
used to implement the link. But typically links will refer to objects
that are not data objects but are contained within a data object.
 The SG knows how to get to the container and the container knows how
to return an element given its index. The job of the link is to find
out the value of the index, persistify it and, later on, pass it on to
the container and get back the linked object. In the next section we
will discuss how links handle indices into generic containers.  

\section{Implementation Techniques}
A big advantage that SG has compared to earlier data
models implementations is that many
compilers are catching up with the ISO/ANSI C++ standard. 
Because of that, a new generation of template libraries like
boost\cite{boost} and loki\cite{Alexandrescu} are bringing once-esoteric
techniques like template meta-programming into the mainstream. 
Template meta-programming uses the compiler template expansion to
control and generate running  code based on static type
information. In SG we have used some of its simpler techniques.
 
\subsection*{Type Traits and Traits Types}
The TDS memory management back-end manages the data objects as
instances of a {\tt DataObject} base class. Each class derived from {\tt DataObject}
has a unique {\tt ClassID}. This allows, for example, to
use an {\it Abstract Factory}\cite{GoF} to create data object
instances when reading from disk. SG wraps each stored data object
into a templated {\tt DataObject}  
\begin{verbatim}
 template <typename DOBJ> 
 class DataBucket : public DataObject {...}
\end{verbatim}   

If {\tt DOBJ} does not inherit from
{\tt DataObject} we want the developer to define
a {\tt ClassID} for {\tt DOBJ} that we will associate to the data object. 

To determine, at compile time, if {\tt DOBJ} inherits from
{\tt DataObject} we use the boost type trait
{\tt boost::is\_base\_and\_derived\brackets{DOBJ,DataObject}}, a
template that evaluates to true when {\tt DOBJ} can be assigned as a
{\tt DataObject}\cite{boost,Alexandrescu}. 

 To associate the {\tt ClassID} information to a data
object type, say vector\brackets{double}, we define a {\tt ClassID\_traits}
structure that developers specialize for that data object (the struct is
actually generated using a cpp macro)
\begin{verbatim} 
 template <> 
 struct ClassID_traits<vector<double> > {
    typedef type_tools::true_tag has_clID_tag;
    static const int ID = 1234;
    ....  
 };
\end{verbatim} 

to manage the {\tt ClassID}s Atlas has developed a simple text-based
``database'' that is used both to generate the{\tt ClassID}s of new
types and to verify at run-time that there are no duplicated {\tt
  ClassID}s and no conflicts.

\subsection*{Concept Checking}
SG allows developers to use generic key types to identify objects of a
given type. A key must of course define an ordering operation. For SG
we also require keys to be persistable. In traditional OO programming
these requirements would be expressed as an interface the key class
imports. In generic programming interfaces are rather exported and
hence verified by the clients. To this end, SG provides a {\tt KeyConcept} 
built using the boost {\tt concept\_check} library  (see
Fig.~\ref{fig:KEYCONCEPT}).
\begin{figure*}
\begin{verbatim} 

  template <typename T, .... > struct KeyConcept {
    void constraints() { 
      boost::function_requires< boost::LessThanComparableConcept<T> >();
      .... 
    }
  };
\end{verbatim} 
  \caption{\label{fig:KEYCONCEPT} Concept Cheching}
\end{figure*}

Inserting in the StoreGate API a call to 
 {\tt boost::function\_requires\brackets{KeyConcept\brackets{KEY}}()}
we allow the compiler to check whether the template parameter {\tt KEY} 
of a retrieve or register method is valid.

\subsection*{Policy Classes}
SG handle and link classes use policy classes to
configure their behavior at compile time. A policy\cite{Alexandrescu} is a statically
configured {\it Strategy}\cite{GoF}. It can also be seen as a traits class
that defines behavior rather than structure. Policies become powerful tools
when they are combined: the compiler picks the right combinations and generates
the code needed by the application. For example the element link class template
{\tt ElementLink} is implemented as a combination of two policies (see
Fig.~\ref{fig:POLICIES}).
\begin{figure*}
\begin{verbatim}

template <typename STORABLE, 
          class StoragePolicy=DataProxyStorage<STORABLE>,
          class IndexingPolicy=typename SG::GenerateIndexingPolicy<STORABLE>::type >
class ElementLink : 
  public StoragePolicy,
  public IndexingPolicy 
{ ... }
\end{verbatim}
  \caption{\label{fig:POLICIES} ElementLink as a combination of policies}
\end{figure*}

{\tt DataProxyStorage} wraps the TDS back-end API,
while {\tt IndexingPolicy} defines the strategy 
the {\tt ElementLink} uses to find a container element given its identifier, 
and viceversa. The type generator template {\tt
  GenerateIndexingPolicy} looks at the data object type ({\tt STORABLE})
and tries to provide a reasonable default strategy for that type.

We have defined indexing policy classes that can be used to
index elements of all STL containers and to index nodes of an HepMC graph\cite{HepMC}. Policies are flexible: if a developer introduces a new
container type, all they have to do is to provide a matching
indexing policy and the compiler will generate the new link type as needed.

\section{Status and Outlook}
After three years of evolution, StoreGate has
achieved a certain maturity. A lot of broad design principles have
been established: work with user types, avoid user-defined keys,
define an access control policy. The core data access API has been 
stable for several releases. The implementation has been reviewed and
reengineered  twice to improve robustness, physical design and
to meet the strict performance requirement of Atlas trigger software\cite{AtlasTrigger}.

In the spirit of the Gaudi open project we have started discussing our
work with the LCG community and we hope the StoreGate ideas and code will be
useful to developers inside and outside ATLAS.

\section*{Acknowledgments}
We would like to thank all ATLAS collaborators who contributed to the design and prototyping of SG. We are extremely grateful to many colleagues from other   
experiments who shared their experiences with us:
M. Frank, V. Innocente, R. Kennedy, J. Kowalkowski, P. Mato,  M. Paterno,
S. Patton, S. Snyder and  L. Tuura.

This work was supported in part by the Office of Science. High Energy Physics ,
U.S. Department of Energy under Contract No. DE-AC03-76SF00098.

\end{document}